\documentclass[aps,reprint,prd]{revtex4-2}
\usepackage{amsmath}
\usepackage{amssymb} 
\usepackage{graphicx}
\usepackage{natbib}
\usepackage{float}
\usepackage[hidelinks]{hyperref}
\usepackage{booktabs}
\usepackage{caption}
\usepackage{subcaption}
\usepackage{gensymb}
\usepackage{listings}
\usepackage{color}
\usepackage{courier}
\usepackage{multirow}
\usepackage{mathtools}

\usepackage[normalem]{ulem}

\usepackage{environ}

\begin{document}
	
	\title{Formation of population inversions in relativistic plasmas through nonresonant interactions with Alfv\'{e}n waves}
	\author{Killian Long}
	\affiliation{University College Cork}

	\author{Asaf Pe'er}
	\affiliation{Bar Ilan University}
 	\begin{abstract}
We solve the full quasilinear kinetic equation governing nonresonant interactions of Alfvén waves with relativistic plasmas. This work was motivated by the need to determine the energy available for the synchrotron maser in the context of Fast Radio Bursts (FRBs). This interaction can result in plasma heating and the formation of population inversions necessary for the maser. We find that population inversions containing $\sim 1-10\%$ of the distribution's energy form in the relativistic regime, providing an explanation for the formation of the inversion in the environment expected near FRBs.
	\end{abstract}
	\date{\today}
	\maketitle
	
The cyclotron or synchrotron maser is a proposed emission mechanism for various astrophysical phenomena such as Auroral Kilometric Radiation (AKR) \cite{1979ApJ...230..621W, Wu1985} in the Earth's magnetosphere, Jovian Decametric Radiation \cite[e.g.][]{1982AuJPh..35..447H} and Fast Radio Bursts (FRBs) \cite[e.g.][]{Lyubarsky2014, 2018ApJ...864L..12L, Metzger2019}, among others. The high brightness temperatures from these objects require a coherent emission mechanism such as synchrotron maser emission (SME) \cite[e.g.][]{2006AaARv..13..229T, 1998JGR...10320159Z, 2016MPLA...3130013K}. Necessary plasma conditions for SME are (i) an inverse population of energetic electrons, and (ii) being embedded in a background magnetic field. Under appropriate conditions, interaction between electromagnetic waves in plasma and the energetic particle population results in negative absorption and stimulated emission \citep{Wu1985,2006AaARv..13..229T}. 

To achieve a population inversion, the particle distribution must either grow faster than $E^2$ \cite{2002ApJ...574..861S}, or satisfy $\frac{\partial F}{\partial v_{\perp}}>0$ \cite{ Wu1985}. Here, $F(v_{\perp},v_{\parallel})$ is the particle distribution function, $v_{\perp} (v_{\parallel})$ is the velocity perpendicular (parallel) to the background magnetic field, and $E$ is the particle's energy. 
Population inversion can be achieved in various ways. For example, in the 'magnetic mirror', particles with larger pitch angles are reflected by the mirror's field while those with small pitch angles escape \cite{1979ApJ...230..621W, 1982JGR....87.5140M}. This leads to an empty cone-shaped void in the distribution in velocity space which satisfies $\frac{\partial F}{\partial v_{\perp}}>0$ \cite{Wu1985, 1976PhFl...19..299F}. Such a scenario was proposed to explain population inversion in the non- or mildly relativistic environments such as AKR \cite[e.g.][]{1979ApJ...230..621W, Wu1985,1982JGR....87.5140M}. Following observations suggesting the 'loss-cone' was insufficient \cite{1990JGR....95.5983L}, the 'ring-shell' model was developed \cite{1985JGR....90.9650P, 1986ApJ...310..432W}, where a combination of the magnetic mirror and  parallel electric fields produces a ring-shaped particle distribution, characterized by the absence of low energy particles. The loss cone is then part of the ring in velocity/momentum space, rather than being part of an isotropic distribution as in the magnetic mirror.  

In contrast to the non- or mildly relativistic plasma in the scenarios discussed above \cite{Wu1985}, the environment around FRB emission regions is expected to be relativistic, namely the particle's temperature is $k_B T \approx m c^2$, where $m$ is the particle mass and $c$ is the speed of light in vacuum \cite{Lyubarsky2021}. Previous works on such environments have focused on inversions formed by strongly magnetised relativistic shocks propagating perpendicularly to the background magnetic field, thereby heating the plasma to relativistic temperatures \cite[e.g.][]{1992ApJ...391...73G, 2006ApJ...653..325A, Plotnikov2019}. At the shock front a soliton-like structure is formed, in which particles gyrate around the enhanced magnetic field and form a semi-coherent ring in momentum space \cite{1988PhFl...31..839A}. Like the distributions discussed above, this distribution can also support SME \cite{1991PhFlB...3..818H}. The existence of this 'ring' has been demonstrated in particle-in-cell (PIC) simulations \cite[e.g.][]{1992ApJ...391...73G, Plotnikov2019, 1988PhFl...31..839A, 2017ApJ...840...52I}.

While these works provide a viable mechanism for the formation of a population inversion, we show here that a population inversion leading to very high efficiency SME can also be naturally achieved through nonresonant interactions between Alfv\'{e}n waves and a plasma without requiring a relativistic shock wave.
In objects such as FRBs the environment is highly magnetized, namely the relativistic plasma is expected to have a low $\beta$, where $\beta$ is the ratio of the plasma and magnetic pressure. This provides the necessary condition that enables the nonresonant interaction between Alfv\'{e}n waves and particles  \cite{Lyubarsky2021}. Due to the problem's complexities, this interaction has previously only been studied in the nonrelativstic regime \cite{Wu2007, Yoon2009}. Nonresonant wave-particle interactions in such plasmas result in pitch-angle diffusion, deforming the initially isotropic distribution into a crescent like shape in momentum space capable of supporting SME, similar to the 'ring-shell' distribution discussed above \cite{Zhao2013, Wu2014}. An example plot of the change in distribution is shown in Figure 1. 

 To investigate the applicability of this mechanism to fully relativistic plasmas, we examine the full nonresonant relativistic interaction with Alfv\'{e}n waves for the first time. Our semi-analytical treatment enables direct measurement of the energy available to the maser as well as the parameter space where the inversion forms. As we show here, high efficiencies of several percent are achieved for conditions that can represent emission from FRBs, providing a direct explanation for the key physical ingredient that underlines their high brightness temperature.

To examine the problem we consider a relativistic plasma of density $n$ embedded in a background magnetic field $\mathbf{B_0}$. The initial particle distribution $F_0(T_0)$ is given by a Maxwell-J\"{u}ttner distribution of temperature $T_0$. As the particles interact with the waves their distribution evolves, and is specified at time $t$ by $F(p_\parallel,p_\perp,t)$, where $p_\perp = \gamma m v_\perp/c$ and $p_{||}=\gamma m v_\parallel/c$ are the relativistic momenta in the perpendicular and parallel directions to the background magnetic field respectively. Here, $\gamma = (1+(p_\perp/m)^2+ (p_\parallel/m)^2)^{1/2} $ is the Lorentz factor. In addition to the steady magnetic field, we assume Alfv\'{e}n waves propagate through, and interact with the plasma. 
These waves 
are taken to have a broad spectrum that only varies slowly in time \cite{Wu2007,Yoon2009}. As the waves are assumed to originate from a central neutron star in the FRB scenario, the wavevector in this case is expected to be in the parallel direction only. We also assume the Alfv\'{e}n waves have a flat spectrum as a test case, though numerical results can be obtained for any spectrum.
The waves are described by spectral magnetic and electric fields of $B_k$ and $E_k$ respectively, where $k$ is the wavenumber. 

We use kinetic theory to study the interaction between the waves and particles. To examine how the particle distribution changes due to the interaction with the Alfv\'{e}n waves, we use the quasilinear approximation of this theory \cite{vedenov1961nonlinear, drummond1962anomalous}, where the variables of the Vlasov equation are split into slowly varying average and first order fluctuation terms. Ref. \cite{stix1992waves} starts from the Vlasov equation, and provides a complete derivation of the full relativistic quasilinear kinetic equation describing the temporal evolution of the particle's distribution function due to interaction with the Alfv\'en waves at resonance. We added nonresonant terms following Ref. \cite{Yoon2009}. The complete equation, that contains both the resonant (line 2) and the nonresonant (line 3) terms and is correct in both relativistic and non-relativistic regimes, is
\begin{eqnarray}
\frac{\partial F}{\partial t}&=&\frac{e^2}{4}\sum_l\int d\textbf{k}\frac{1}{p_{\perp}}\left[\left(1-\frac{k_\parallel p_\parallel}{\gamma m\omega}\right)\frac{\partial}{\partial p_\perp}+\frac{k_\parallel p_\perp}{\gamma m\omega}\frac{\partial}{\partial p_\parallel}\right]\nonumber\\&&\times\Bigg\lbrace p_\perp\Bigg[\pi\delta\left(\omega-l\omega_c-\frac{k_\parallel p_\parallel}{\gamma m}\right)|E_k|^2\nonumber\\&&-\frac{\partial}{2\partial\omega}\left(PV\left(\frac{1}{\omega-l\omega_c-\frac{k_\parallel p_\parallel}{\gamma m}}\right)\right)\frac{\partial\left|E_k\right|^2}{\partial t}\Bigg]\nonumber\\&&\times\left[\left(1-\frac{k_\parallel p_\parallel}{\gamma m\omega}\right)\frac{\partial}{\partial p_\perp}+\frac{k_\parallel p_\perp}{\gamma m\omega}\frac{\partial}{\partial p_\parallel}\right]F\Bigg\rbrace,
\label{eq:2}
\end{eqnarray} 
 Here $e$ is the unit electric charge; $\omega_c = eB_0/\gamma mc$ is the cyclotron frequency; $\omega$ is the Alfv\'{e}n wave frequency given by the linear theory; $PV$ is the Principal Value; $k_\parallel$ is the wavenumber in the direction parallel to the background magnetic field, and $l=1,2,...$ is the harmonic number. 

The particle distribution's evolution is determined by the partial derivative terms and their coefficients. First and second order derivatives in both directions are present, which describe advection and diffusion respectively. The coefficients' relative magnitude dictates which process is most influential. These depend on the linear Alfv\'{e}n dispersion relation $\omega(k)$, as each contains the factor $(k_\parallel/\omega)^\alpha$, where $\alpha = 0,1,2$. The equation is also governed by whether the resonance condition $\omega -l\omega_c - \frac{k_\parallel p_\parallel}{\gamma m}=0$ is satisfied. While this is the case in many plasmas, it is not in FRB emission regions. Due to strong magnetic fields, plasmas in these environments have a low $\beta$. For Alfv\'{e}n waves in such a plasma, the inequalities $\omega_c \gg \omega$ and  $\omega_c \gg \frac{k_\parallel p_\parallel}{\gamma m}$ hold \cite{Wu2007}, 
resulting in no contribution from the resonant term (note that when the nonresonant term is dominant the evolution depends on the temporal change of the Alfv\'{e}n wave electric field $\frac{\partial |E_k|^2}{\partial t}$, rather than its magnitude as in the resonant case).

As for the non-resonant term, using the inequalities above, the Principal Value part of the equation can be simplified as follows \cite{Yoon2009}:
\begin{eqnarray}
\frac{\partial}{2\partial\omega}\left(PV\left(\frac{1}{\omega-l\omega_c-\frac{k_\parallel p_\parallel}{\gamma m}}\right)\right)\frac{\partial\left|E_k\right|^2}{\partial t}&\approx&-\frac{1}{2\omega_c^2}\frac{\partial\left|E_k\right|^2}{\partial t},
\end{eqnarray}
where the higher harmonics' ($l=2,3,..$) contribution is neglected. 

After expanding the equation and separating the terms for clarity, using only the dominant ($l=1$) harmonic, Equation \ref{eq:2} becomes 
\begin{eqnarray}
\frac{\partial F}{\partial t}&=&\frac{c}{8B_0^2}\Bigg\lbrace \left(2\frac{I_1}{  \gamma}-2\frac{I_2q_\parallel}{ \gamma^2}\right)\frac{\partial F}{\partial q_\parallel}\nonumber \\
&&+\left(\frac{1}{q_{\perp}}\left(I_3+\frac{I_2q_\parallel^2}{\gamma^2 }-2\frac{I_1 q_\parallel}{\gamma }\right)-\frac{I_2q_\perp}{\gamma^2}\right)\frac{\partial F}{\partial q_\perp}\nonumber\\
&&+\frac{I_2 q_\perp^2}{\gamma^2 }\frac{\partial^2 F}{\partial q_\parallel^2}\nonumber\\
&&+\left(I_3+\frac{I_2q_\parallel^2}{\gamma^2}-2\frac{I_1 q_\parallel}{\gamma }\right)\frac{\partial^2 F}{\partial q_\perp^2}\nonumber\\
&&+\left(2\frac{ I_1 q_\perp}{\gamma }-2\frac{I_2 q_\perp q_\parallel}{\gamma^2 }\right)\frac{\partial^2 F}{\partial q_\parallel q_\perp}\Bigg\rbrace,
\label{eq:4}
\end{eqnarray} 
where $q=p/m=\gamma \frac{v}{c}=\gamma\beta$, and $I_1 = \int d\textbf{k}\frac{\partial\left|E_k\right|^2}{\partial t}\frac{k_\parallel}{\omega}$, $I_2 = \int d\textbf{k}\frac{\partial\left|E_k\right|^2}{\partial t}\frac{ck_\parallel^2}{\omega^2}$ and $I_3 = \int d\textbf{k}\frac{\partial\left|E_k\right|^2}{\partial t}\frac{1}{c}$. Equation \ref{eq:4} is correct in the limit of strong magnetic field, in both the relativistic and non-relativistic regimes. 
The integrals $I_1$ and $I_2$ are calculated using either the nonrelativistic Alfv\'{e}n dispersion relation $\omega =kv_A$, where $v_A$ is the Alfv\'{e}n velocity, or the full relativistic solution given by \cite{Asenjo2009,Munoz}
\begin{equation}
\omega^2-c^2k^2=\underset{s}{\sum}\omega_{p(s)}^2\frac{\omega'}{f_s\gamma_s \omega'-\omega_{cs}},
\end{equation}
where $\omega'=\omega - kv_0$; $v_0$ is the particles' bulk velocity (in case it is non-zero); $\omega_{p(s)} =\sqrt{4\pi n_s e^2/m_s}$ is the plasma frequency; $f_s=K_3\left(\frac{m_sc^2}{k_B  T_s} \right)\Big/K_2\left(\frac{m_s c^2}{k_B T_s}\right)$ is the plasma's enthalpy; and the subscript $s$ denotes the particle species (electrons and ions). Here $K_l$ is a modified Bessel function of the second kind of order $l$. In the nonrelativistic regime the ratio between the integrals is simply given by $I_1 = \frac{v_A}{c}I_2 =\frac{c}{v_A}I_3$. In this limit therefore $I_2>I_1>I_3$ for all $v_A<c$, though such a concise connection does not exist for the relativistic case.

We solve equation \ref{eq:4} numerically using MATLAB on a term by term basis using a combination of the Crank-Nicolson and upwind differencing finite difference methods \cite{Press2007}. The initial conditions are the initial distribution function $F_0$; which is taken to be a Maxwell-J\"{u}ttner distribution with a normalized temperature $\theta =\frac{k_BT}{mc^2}$; the magnetisation $\sigma = \Omega^2/\omega_p^2$; as well as $k_{max}$ and $k_{min}$. Here, $\Omega = eB_0/mc$ is the gyration frequency of the plasma as a whole. We further assume a single temperature and density describing all species. To check the code's accuracy, comparisons were made to the analytical nonrelativistic solution found in Equation 9 of Ref. \cite{Wu2007}.  In all cases investigated the full solution and nonrelativistic result agree \footnote{See Supplemental Material at [URL will be inserted by publisher] for figure showing example comparisons}.
\begin{figure}
	\centering
	\captionsetup{width=\columnwidth}
	
	\includegraphics[width=1\columnwidth]{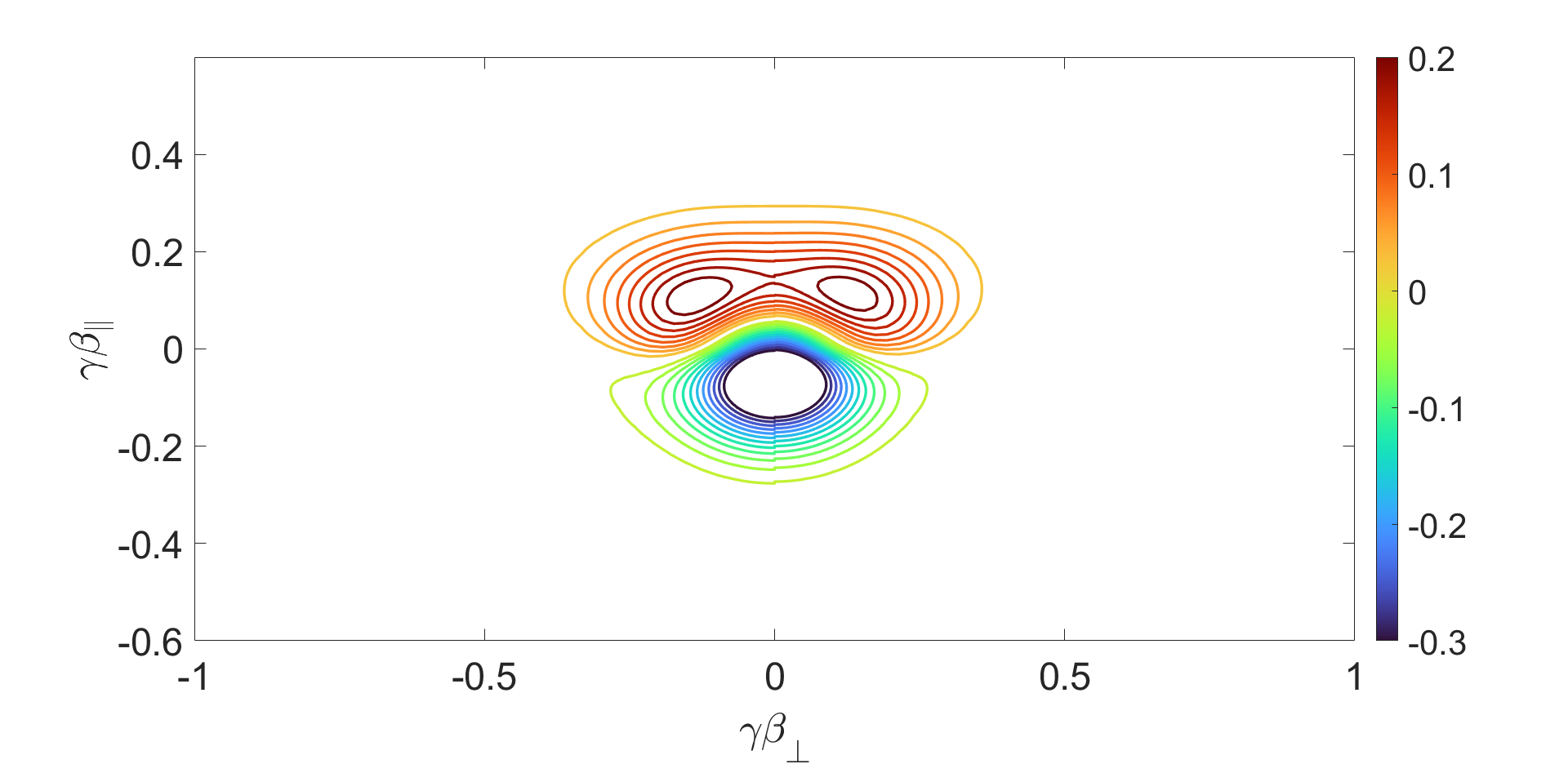}
	\caption{A contour plot showing the difference between the particle distribution at $\Omega \tau = 0.23$ compared to an initial Maxwell-J\"{u}ttner distribution with $\theta=0.01$ and $\sigma = 0.1$.
		 The formation of the crescent shape is observed by noting that the increase in parallel momentum is larger closer to $\gamma \beta_\perp=0$.}
	\label{fig:2}
\end{figure}

 We present results for four scenarios with temperatures $\theta = 0.01$, $\theta = 0.1$ and $\theta=1$ as evidence for the formation of population inversions in relativistic plasmas. These runs show the change from the initial symmetric Maxwell-J\"{u}ttner distribution to the crescent shape required for the inversion. The time units are given in terms of $\Omega\tau$. Here $\tau = \Gamma\eta t$, where $\Gamma<<1$ is the temporal growth rate of the Alfv\'{e}n wave relative to $\Omega$, and $\eta = |E_k|^2/B_0^2<<1$. Figure 1 shows a contour plot of the difference between the particle distribution at $\Omega \tau = 100$, and an initial Maxwell-J\"{u}ttner distribution with $\theta=0.01$, $\sigma=0.1$, $k_{min}=0$ and $k_{max}=0.1\Omega/c$. The distribution evolves primarily in the parallel direction through advection, which has two characteristics which result in a population inversion. Firstly, for certain parameters, the advection is in the same direction for all values of $q_\parallel$, resulting in motion in the positive parallel direction only. This can be seen by examining the coefficient of the $\partial F/\partial q_\parallel$ term in Eq. \ref{eq:4}. This term's sign changes at $q_\parallel = \frac{I_1\sqrt{1+q_\perp^2}}{\sqrt{I_2^2-I_1^2}}$. Therefore, when $I_2<I_1$, no sign change ever occurs, while for $I_2>I_1$ the value of $q_\parallel$ at which the sign changes becomes closer to zero as $I_2/I_1$ increases. If $I_2/I_1$ is too large, no crescent shape can form as the direction of motion will change near $q_\parallel=0$. Secondly, the advection depends on $q_\perp$, resulting in varying advection strength across the distribution.  Perpendicularly, there is only very slight diffusion, with no advection. Together, these features result in deformation into the expected crescent shape, with the distribution near $q_\perp=0$ advecting further in the positive parallel direction than the distribution at larger values of $q_\perp$. This can be seen in Figure 1, with the largest change in the distribution centred at $q_\parallel = 0.14$ for $q_\perp = 0$, compared to the smaller value of $q_\parallel = 0.09$ for $q_\perp = 0.2$. Similar behaviour is found in the distributions with higher initial temperatures, though on longer timescales.
%
\begin{figure}
	\centering
	
	\captionsetup{width=\columnwidth,justification=justified}

	\includegraphics[width=\columnwidth]{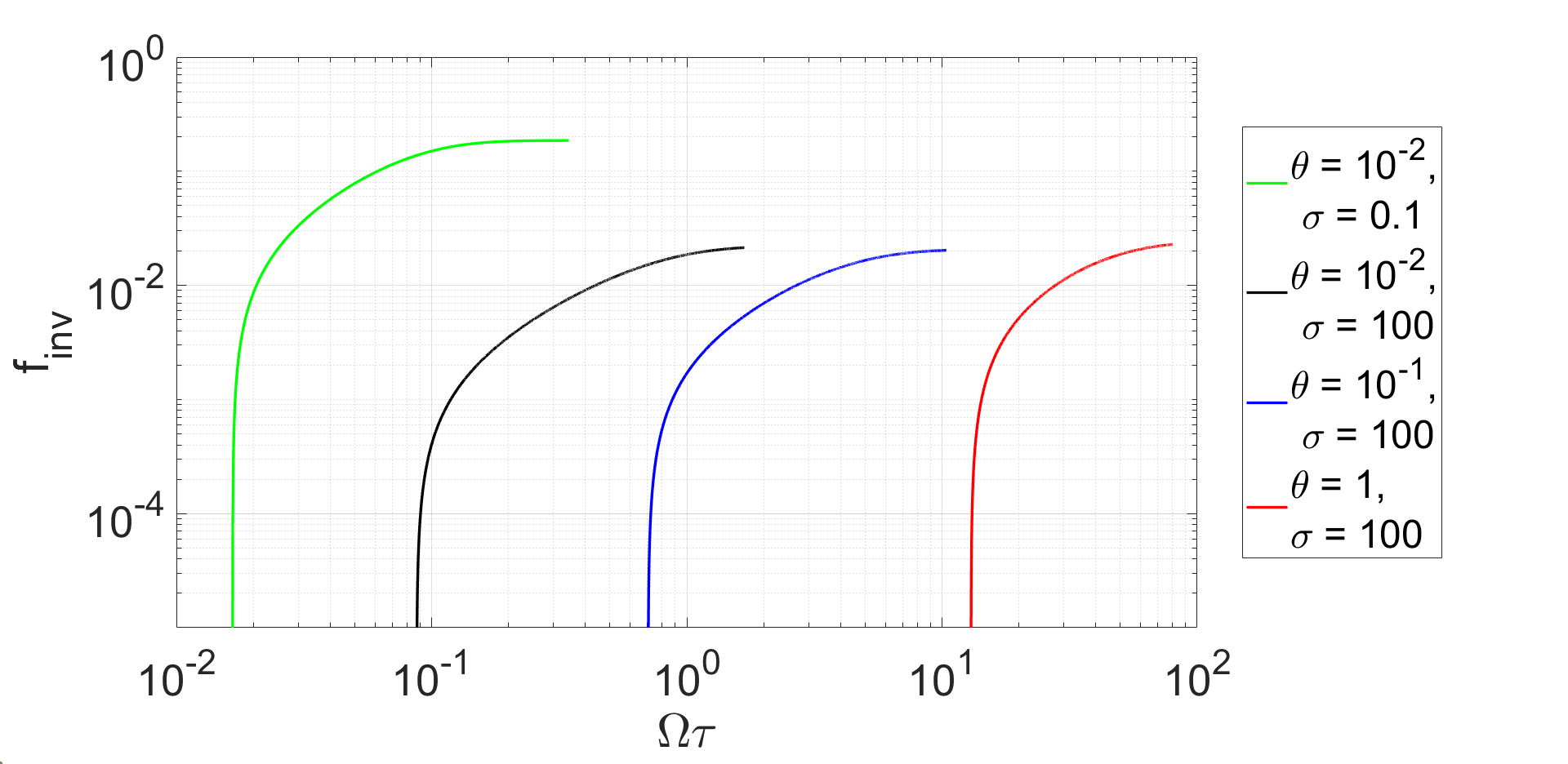}
	\caption{Fraction of energy in the inversion ($f_{inv}$) versus time for initial Maxwell-J\"{u}ttner distributions with $\theta=0.01,\,\sigma = 0.1$ (green),   $\theta=0.01,\,\sigma = 100$ (black), $\theta=0.1,\,\sigma = 100$ (blue) and  $\theta=1,\,\sigma = 100$ (red). $k_{min}$ = 0, and $k_{max} = 0.1\Omega/c$ in all four cases.  }
	\label{fig:3}
\end{figure}

To determine the fraction of energy in the crescent-shaped population inversion, $f_{inv}$, we compare the energy of the distribution, $E_{tot}$, to the total energy of a bimaxwellian distribution, $E_{bi}$, with the same $\theta_{\|}$ and $\theta_{\perp}$. $f_{inv}$ is then obtained using $f_{inv} = (E_{tot}-E_{bi})/E_{tot}$, where $E_{tot}$ is the energy of the actual distribution. $f_{inv}$ therefore is the amount of excess energy contained in the crescent of the deformed distribution, but does not include any contributions from changes in energy that are not associated with the population inversion. The value of $f_{inv}$ is therefore the absolute upper limit on the amount of energy available for masing.

 The fraction of energy contained in the inversion $f_{inv}$  increases with time for all four parameters sets, as seen in Fig. \ref{fig:3}. This figure shows $f_{inv}$ for $\lbrace\theta,\sigma\rbrace$ of $\lbrace10^{-2},0.1\rbrace$, $\lbrace10^{-2},100\rbrace$, $\lbrace10^{-1},100\rbrace$ and $\lbrace1,100\rbrace$. For all four cases, $k_{min} = 0$ and $k_{max} = 0.1\Omega/c$.
  In all cases the trend is the same. Once the inversion begins to form $f_{inv}$ initially increases steeply, and continues to increase with gradually decreasing rates of change at longer times, before converging to a value of $f_{inv} \sim 0.19$ at $\Omega\tau > 0.2 $ in the case of $\lbrace\theta = 10^{-2},\sigma = 0.1\rbrace$, $f_{inv} \sim 0.02$ at $\Omega\tau > 1.6$ for $\lbrace 10^{-2}, 100\rbrace$, $f_{inv} \sim 0.02$ at $\Omega\tau > 10$ for $\lbrace10^{-1},100\rbrace$ and $f_{inv} \sim 0.02$ at $\Omega\tau > 80$ for $\lbrace1,100\rbrace$. While the asymptotic values of $f_{inv}$ are similar for the different temperatures, the timescales become longer as $\theta$ increases.

It is useful to compare the efficiency in the population inversion achieved by non-resonant interactions, to that found by numerical studies of relativistic shocks, as both are viable candidates in explaining SME in relativistic environments \cite[e.g.][]{1992ApJ...391...73G, 2006ApJ...653..325A, Plotnikov2019}. For $\sigma >> 1$, an efficiency of $\sim10^{-3}\sigma^{-1}$ was found in relativistic shocks \cite{Sironi2021}. For $\sigma = 100$ this is an efficiency of $\sim10^{-5}$, three orders of magnitude less than the values found in this work. However, for $\sigma\sim0.1$ the shock efficiency was found to be $\sim 0.1$ \cite{Plotnikov2019}, the same order as the asymptotic value of $f_{inv} \sim 0.19$ found in this work.  It is important to note that the efficiency in the relativistic shock scenario and $f_{inv}$ is not a one-to-one comparison, as while the efficiency provides a direct measure of the amount of energy radiated by the maser, $f_{inv}$ on the other hand is an upper limit on the total energy available, which may not be fully extracted due to the influence of other plasma processes. For instance, the SME growth rate may be reduced at higher temperatures \cite{2006ApJ...653..325A, 2017ApJ...840...52I, 2020MNRAS.499.2884B}, potentially reducing the amount of energy extracted due to the impact of other plasma instabilities.

 To estimate the importance of these instabilities, we conduct the following brief analysis. We note that the deformed distribution is close to a bimaxwellian, which is susceptible to the firehose and mirror instabilities. These have stability criteria of $T_\perp/T_\parallel > 1-1/\beta_\parallel$ and $T_\perp/T_\parallel < 1+1/\beta_\parallel$ respectively \cite{gary1993theory}. Here $\beta_\parallel = 2 \theta_\parallel/\sigma$ is the parallel plasma beta. For $\sigma = 100$, $\beta_\parallel$ is therefore always small. For the $\sigma = 0.1$ case, the low temperature ensures $\beta_\parallel<<1$. Combined with the small temperature anisotropies ($0.5<T_\perp/T_\parallel< 1$), both criteria are satisfied for the runs presented. As the distribution is not susceptible to these other plasma instabilities for these parameters, the maser should therefore extract a significant fraction of $f_{inv}$, provided that the growth rate is sufficiently fast to extract the energy within the size of the masing cavity.
 	
 We do point out that this will not be the case for some parameters not explored in this work, especially those with $\sigma < 1$ and higher temperatures, where the growth rates of the mirror and firehose instabilities can become significant. In such a scenario, the SME energy may be lower than $f_{inv}$. The plasma may also be susceptible to turbulent cascades when Alfv\'{e}n wave spectra with perpendicular components are considered \cite{2022PhRvL.128g5101N}.

In order for this mechanism to be applicable to FRBs the inversion must form with realistic parameter values. Observed signals from FRBs are in the GHz band, with frequencies of 111 MHz \cite{2019ARep...63...39F} to 8 GHz \cite{2018ApJ...863....2G}.  When $\sigma>1$, the  maser's peak frequency is $\omega_m\sim\Omega\gamma^{-1}$ \cite{Lyubarsky2021}, requiring magnetic fields of $B_0 \sim (40-2.9\times10^3)\gamma \text{ G}$ and maximum number densities of $n\sim(1.5\times10^8-7.9\times 10^{11})\gamma\text{ cm}^{-3}$. When $\sigma<1$, SME peaks at  $\omega_m\approx\omega_p \text{min}\left\lbrace\gamma,\sigma^{-1/4}\right\rbrace$ \cite{2002ApJ...574..861S}. Due to the very weak dependence on $\sigma$, the peak frequency will never be much greater than $\omega_p$, allowing the approximation $\omega_m \sim \omega_p$. This results in a requirement of $n\sim(1.5\times10^8-7.9\times 10^{11})\gamma\text{ cm}^{-3}$, with a maximum allowed background field of $B_0 \sim(40-2.9\times 10^3)\gamma\text{ G}$, using $\sigma<1$. As the values for the magnetic field in the $\sigma < 1$ case and the number density in the $\sigma>1$ case are both upper limits, a large parameter space is viable for SME.

The population inversion formation time must also be realistic. The leading candidates for FRB progenitors are magnetars, neutron stars with extremely high surface magnetic fields of $\sim 10^{15}$ G, with the galactic magnetar SGR 1935+2154 directly associated with an FRB \cite{2020Natur.587...54C}. Magnetars have two timescales of interest to the variability in the Alfv\'{e}n wave field necessary for nonresnonant wave particle interaction. Firstly, the magnetar radius $R_*$ corresponds to a timescale of $\sim R_*/c = 3.3\times 10^{-5}R_{*,6}\text{ s}$. Here, $R_{*,x}=R_*/10^{x}$. The second timescale is the star's period, typically $\sim 1-10\text{ s}$ \cite{2008A&ARv..15..225M, 2014ApJS..212....6O}, \footnote{http://www.physics.mcgill.ca/~pulsar/magnetar/main.html},
 though young magnetars are expected to have millisecond periods \cite{2022ASSL..465..245D}.
The time taken to reach the asymptotic timescale is $t\sim2\times10^{-10}(\Omega_9 \Gamma \eta)^{-1}\,\text{s}$ for $\lbrace\theta =0.01,\sigma=0.1\rbrace$, $t\sim1.6\times10^{-9}(\Omega_9 \Gamma \eta)^{-1}\,\text{s}$ for $\lbrace0.01,100\rbrace$, $t\sim10^{-8}(\Omega_9 \Gamma \eta)^{-1}\,\text{s}$ for $\lbrace0.1,100\rbrace$,  and $t\sim8\times10^{-8}(\Omega_9 \Gamma \eta)^{-1}\,\text{s}$ for $\lbrace1,100\rbrace$. Therefore, the neutron star period timescale is attainable for values of $\Gamma \eta > 2\times10^{-11}\Omega_9^{-1}$, $\Gamma \eta > 1.6\times10^{-10}\Omega_9^{-1}$, $\Gamma \eta > 10^{-9}\Omega_9^{-1}$ and $\Gamma \eta > 8\times10^{-9}\Omega_9^{-1}$ respectively. However, the neutron star radius timescale imposes considerably more stringent limits of $\Gamma \eta > 6\times10^{-6}(\Omega_9R_{*,6})^{-1}$ for $\lbrace0.01,0.1\rbrace$, $\Gamma \eta > 5\times10^{-5}(\Omega_9R_{*,6})^{-1}$ for $\lbrace0.01,100\rbrace$, $\Gamma \eta > 3\times10^{-4}(\Omega_9R_{*,6})^{-1}$ for $\lbrace0.1,100\rbrace$ and $\Gamma \eta > 2.4\times10^{-3}(\Omega_9R_{*,6})^{-1}$ for $\lbrace1,100\rbrace$. As both $\Gamma<<1$ and $\eta<<1$, these results show that higher temperatures may not be viable on this timescale.

The density and magnetic field constraints derived above can be satisfied in two different magnetar scenarios.  Firstly, we consider a relativistic magnetar wind, where the magnetisation is comparable to this work, with $  \sigma_{wind}\sim280B_{*,15}^{8/9}R_{*,6}^{24/9}M_3^{-2/3}P^{-2}
$ \cite{Lyubarsky2021}. Here $B_*$ is the surface magnetic field, $M$ is the pair multiplicity, and $P$ is the period. Using the required range for $B_0$ when $\sigma>1$, emission could take place at distances of $R\sim(1.5\times 10^{10}-1.1\times10^{12})\gamma^{-1}B_{*,15}R_{*,6}^3P^{-2}\text{ cm}$ from the neutron star, assuming the azimuthal magnetic field $B\propto1/R$ dominates outside the light cylinder. The number density at this distance is significantly below the upper limits, indicating that SME is possible in magnetar winds. 

Another possible scenario involves a subrelativistic wind with $\beta_w\lesssim 1$ and $\sigma_w\lesssim 1$ from a magnetar or other compact object \cite{Metzger2019}. In this case the wind density is $   n \sim 1.6\times10^{5}\dot{\mathcal{M}}_{21}R_{14}^{-2}\beta_w^{-1}\text{ cm}^{-3}$. Here, $\dot{\mathcal{M}}$ is the mass loss rate, which is normalised to constraints obtained from FRB 121102, where $\dot{\mathcal{M}}\sim10^{19}-10^{21}\text{ g s}^{-1}$ \cite{2018ApJ...868L...4M}. At distances of greater than $R\sim 4.5\times 10^{10}-3.3\times10^{12}\text{ cm}$ this environment satisfies the constraints for $n$ and $B$ in the $\sigma < 1 $ regime, providing another region in which the population inversion could form. 

The results presented in this Letter show that nonresonant interaction of Alfv\'{e}n waves with a relativistic plasma can produce population inversions for temperatures in the range $\theta = 0.01-1$, and that a large fraction of the overall energy is contained in the inversion, with values of $f_{inv}\geq0.01$. This energy is comparable to or greater than that expected from SME from relativistic shocks in similar conditions. Furthermore, we have demonstrated that these conditions are realistic for two models of FRBs. A further exploration of the full parameter space in temperature, magnetisation and wavenumber range is currently in preparation, and will be presented in a future work.

 AP acknowledges support from the European Union (EU) via ERC consolidator grant 773062 (O.M.J.). KL acknowledges the support of the Irish Research Council through grant number GOIPG/2017/1146 as well as funding obtained by the above ERC grant. The authors also  wish to acknowledge the Irish Centre for High-End Computing (ICHEC) for the provision of computational facilities and support. We would also like to thank the referees for their many useful comments and Antoine Bret for helpful discussions.

\bibliographystyle{apsrev4-2}
\bibliography{paper_fin}

\begin{thebibliography}{45}%
\makeatletter
\providecommand \@ifxundefined [1]{%
 \@ifx{#1\undefined}
}%
\providecommand \@ifnum [1]{%
 \ifnum #1\expandafter \@firstoftwo
 \else \expandafter \@secondoftwo
 \fi
}%
\providecommand \@ifx [1]{%
 \ifx #1\expandafter \@firstoftwo
 \else \expandafter \@secondoftwo
 \fi
}%
\providecommand \natexlab [1]{#1}%
\providecommand \enquote  [1]{``#1''}%
\providecommand \bibnamefont  [1]{#1}%
\providecommand \bibfnamefont [1]{#1}%
\providecommand \citenamefont [1]{#1}%
\providecommand \href@noop [0]{\@secondoftwo}%
\providecommand \href [0]{\begingroup \@sanitize@url \@href}%
\providecommand \@href[1]{\@@startlink{#1}\@@href}%
\providecommand \@@href[1]{\endgroup#1\@@endlink}%
\providecommand \@sanitize@url [0]{\catcode `\\12\catcode `\$12\catcode
  `\&12\catcode `\#12\catcode `\^12\catcode `\_12\catcode `\%12\relax}%
\providecommand \@@startlink[1]{}%
\providecommand \@@endlink[0]{}%
\providecommand \url  [0]{\begingroup\@sanitize@url \@url }%
\providecommand \@url [1]{\endgroup\@href {#1}{\urlprefix }}%
\providecommand \urlprefix  [0]{URL }%
\providecommand \Eprint [0]{\href }%
\providecommand \doibase [0]{https://doi.org/}%
\providecommand \selectlanguage [0]{\@gobble}%
\providecommand \bibinfo  [0]{\@secondoftwo}%
\providecommand \bibfield  [0]{\@secondoftwo}%
\providecommand \translation [1]{[#1]}%
\providecommand \BibitemOpen [0]{}%
\providecommand \bibitemStop [0]{}%
\providecommand \bibitemNoStop [0]{.\EOS\space}%
\providecommand \EOS [0]{\spacefactor3000\relax}%
\providecommand \BibitemShut  [1]{\csname bibitem#1\endcsname}%
\let\auto@bib@innerbib\@empty
\bibitem [{\citenamefont {{Wu}}\ and\ \citenamefont
  {{Lee}}(1979)}]{1979ApJ...230..621W}%
  \BibitemOpen
  \bibfield  {author} {\bibinfo {author} {\bibfnamefont {C.~S.}\ \bibnamefont
  {{Wu}}}\ and\ \bibinfo {author} {\bibfnamefont {L.~C.}\ \bibnamefont
  {{Lee}}},\ }\href {https://doi.org/10.1086/157120} {\bibfield  {journal}
  {\bibinfo  {journal} {Astrophysical Journal}\ }\textbf {\bibinfo {volume}
  {230}},\ \bibinfo {pages} {621} (\bibinfo {year} {1979})}\BibitemShut
  {NoStop}%
\bibitem [{\citenamefont {Wu}(1985)}]{Wu1985}%
  \BibitemOpen
  \bibfield  {author} {\bibinfo {author} {\bibfnamefont {C.~S.}\ \bibnamefont
  {Wu}},\ }\bibfield  {journal} {\bibinfo  {journal} {Space Science Reviews}\
  }\textbf {\bibinfo {volume} {41}},\ \href
  {https://doi.org/10.1007/BF00190653} {10.1007/BF00190653} (\bibinfo {year}
  {1985})\BibitemShut {NoStop}%
\bibitem [{\citenamefont {{Hewitt}}\ \emph {et~al.}(1982)\citenamefont
  {{Hewitt}}, \citenamefont {{Melrose}},\ and\ \citenamefont
  {{Ronnmark}}}]{1982AuJPh..35..447H}%
  \BibitemOpen
  \bibfield  {author} {\bibinfo {author} {\bibfnamefont {R.~G.}\ \bibnamefont
  {{Hewitt}}}, \bibinfo {author} {\bibfnamefont {D.~B.}\ \bibnamefont
  {{Melrose}}},\ and\ \bibinfo {author} {\bibfnamefont {K.~G.}\ \bibnamefont
  {{Ronnmark}}},\ }\href {https://doi.org/10.1071/PH820447} {\bibfield
  {journal} {\bibinfo  {journal} {Australian Journal of Physics}\ }\textbf
  {\bibinfo {volume} {35}},\ \bibinfo {pages} {447} (\bibinfo {year}
  {1982})}\BibitemShut {NoStop}%
\bibitem [{\citenamefont {Lyubarsky}(2014)}]{Lyubarsky2014}%
  \BibitemOpen
  \bibfield  {author} {\bibinfo {author} {\bibfnamefont {Y.}~\bibnamefont
  {Lyubarsky}},\ }\bibfield  {journal} {\bibinfo  {journal} {Monthly Notices of
  the Royal Astronomical Society: Letters}\ }\textbf {\bibinfo {volume}
  {442}},\ \href {https://doi.org/10.1093/mnrasl/slu046}
  {10.1093/mnrasl/slu046} (\bibinfo {year} {2014})\BibitemShut {NoStop}%
\bibitem [{\citenamefont {{Long}}\ and\ \citenamefont
  {{Pe'er}}(2018)}]{2018ApJ...864L..12L}%
  \BibitemOpen
  \bibfield  {author} {\bibinfo {author} {\bibfnamefont {K.}~\bibnamefont
  {{Long}}}\ and\ \bibinfo {author} {\bibfnamefont {A.}~\bibnamefont
  {{Pe'er}}},\ }\href {https://doi.org/10.3847/2041-8213/aada0b} {\bibfield
  {journal} {\bibinfo  {journal} {Astrophysical Journal Letters}\ }\textbf
  {\bibinfo {volume} {864}},\ \bibinfo {eid} {L12} (\bibinfo {year} {2018})},\
  \Eprint {https://arxiv.org/abs/1806.02700} {arXiv:1806.02700 [astro-ph.HE]}
  \BibitemShut {NoStop}%
\bibitem [{\citenamefont {Metzger}\ \emph {et~al.}(2019)\citenamefont
  {Metzger}, \citenamefont {Margalit},\ and\ \citenamefont
  {Sironi}}]{Metzger2019}%
  \BibitemOpen
  \bibfield  {author} {\bibinfo {author} {\bibfnamefont {B.~D.}\ \bibnamefont
  {Metzger}}, \bibinfo {author} {\bibfnamefont {B.}~\bibnamefont {Margalit}},\
  and\ \bibinfo {author} {\bibfnamefont {L.}~\bibnamefont {Sironi}},\
  }\bibfield  {journal} {\bibinfo  {journal} {Monthly Notices of the Royal
  Astronomical Society}\ }\textbf {\bibinfo {volume} {485}},\ \href
  {https://doi.org/10.1093/mnras/stz700} {10.1093/mnras/stz700} (\bibinfo
  {year} {2019})\BibitemShut {NoStop}%
\bibitem [{\citenamefont {{Treumann}}(2006)}]{2006AaARv..13..229T}%
  \BibitemOpen
  \bibfield  {author} {\bibinfo {author} {\bibfnamefont {R.~A.}\ \bibnamefont
  {{Treumann}}},\ }\href {https://doi.org/10.1007/s00159-006-0001-y} {\bibfield
   {journal} {\bibinfo  {journal} {Astronomy and Astrophysics Review}\ }\textbf
  {\bibinfo {volume} {13}},\ \bibinfo {pages} {229} (\bibinfo {year}
  {2006})}\BibitemShut {NoStop}%
\bibitem [{\citenamefont {{Zarka}}(1998)}]{1998JGR...10320159Z}%
  \BibitemOpen
  \bibfield  {author} {\bibinfo {author} {\bibfnamefont {P.}~\bibnamefont
  {{Zarka}}},\ }\href {https://doi.org/10.1029/98JE01323} {\bibfield  {journal}
  {\bibinfo  {journal} {Journal of Geophysical Research}\ }\textbf {\bibinfo
  {volume} {103}},\ \bibinfo {pages} {20159} (\bibinfo {year}
  {1998})}\BibitemShut {NoStop}%
\bibitem [{\citenamefont {{Katz}}(2016)}]{2016MPLA...3130013K}%
  \BibitemOpen
  \bibfield  {author} {\bibinfo {author} {\bibfnamefont {J.~I.}\ \bibnamefont
  {{Katz}}},\ }\href {https://doi.org/10.1142/S0217732316300135} {\bibfield
  {journal} {\bibinfo  {journal} {Modern Physics Letters A}\ }\textbf {\bibinfo
  {volume} {31}},\ \bibinfo {eid} {1630013} (\bibinfo {year} {2016})},\ \Eprint
  {https://arxiv.org/abs/1604.01799} {arXiv:1604.01799 [astro-ph.HE]}
  \BibitemShut {NoStop}%
\bibitem [{\citenamefont {{Sagiv}}\ and\ \citenamefont
  {{Waxman}}(2002)}]{2002ApJ...574..861S}%
  \BibitemOpen
  \bibfield  {author} {\bibinfo {author} {\bibfnamefont {A.}~\bibnamefont
  {{Sagiv}}}\ and\ \bibinfo {author} {\bibfnamefont {E.}~\bibnamefont
  {{Waxman}}},\ }\href {https://doi.org/10.1086/340948} {\bibfield  {journal}
  {\bibinfo  {journal} {\apj}\ }\textbf {\bibinfo {volume} {574}},\ \bibinfo
  {pages} {861} (\bibinfo {year} {2002})},\ \Eprint
  {https://arxiv.org/abs/astro-ph/0202337} {arXiv:astro-ph/0202337 [astro-ph]}
  \BibitemShut {NoStop}%
\bibitem [{\citenamefont {{Melrose}}\ \emph {et~al.}(1982)\citenamefont
  {{Melrose}}, \citenamefont {{R{\"o}nnmark}},\ and\ \citenamefont
  {{Hewitt}}}]{1982JGR....87.5140M}%
  \BibitemOpen
  \bibfield  {author} {\bibinfo {author} {\bibfnamefont {D.~B.}\ \bibnamefont
  {{Melrose}}}, \bibinfo {author} {\bibfnamefont {K.~G.}\ \bibnamefont
  {{R{\"o}nnmark}}},\ and\ \bibinfo {author} {\bibfnamefont {R.~G.}\
  \bibnamefont {{Hewitt}}},\ }\href {https://doi.org/10.1029/JA087iA07p05140}
  {\bibfield  {journal} {\bibinfo  {journal} {Journal of Geophysical Research}\
  }\textbf {\bibinfo {volume} {87}},\ \bibinfo {pages} {5140} (\bibinfo {year}
  {1982})}\BibitemShut {NoStop}%
\bibitem [{\citenamefont {{Freund}}\ and\ \citenamefont
  {{Wu}}(1976)}]{1976PhFl...19..299F}%
  \BibitemOpen
  \bibfield  {author} {\bibinfo {author} {\bibfnamefont {H.~P.}\ \bibnamefont
  {{Freund}}}\ and\ \bibinfo {author} {\bibfnamefont {C.~S.}\ \bibnamefont
  {{Wu}}},\ }\href {https://doi.org/10.1063/1.861440} {\bibfield  {journal}
  {\bibinfo  {journal} {Physics of Fluids}\ }\textbf {\bibinfo {volume} {19}},\
  \bibinfo {pages} {299} (\bibinfo {year} {1976})}\BibitemShut {NoStop}%
\bibitem [{\citenamefont {{Louarn}}\ \emph {et~al.}(1990)\citenamefont
  {{Louarn}}, \citenamefont {{Roux}}, \citenamefont {{de F{\'e}raudy}},
  \citenamefont {{Le Qu{\'e}au}}, \citenamefont {{Andr{\'e}}},\ and\
  \citenamefont {{Matson}}}]{1990JGR....95.5983L}%
  \BibitemOpen
  \bibfield  {author} {\bibinfo {author} {\bibfnamefont {P.}~\bibnamefont
  {{Louarn}}}, \bibinfo {author} {\bibfnamefont {A.}~\bibnamefont {{Roux}}},
  \bibinfo {author} {\bibfnamefont {H.}~\bibnamefont {{de F{\'e}raudy}}},
  \bibinfo {author} {\bibfnamefont {D.}~\bibnamefont {{Le Qu{\'e}au}}},
  \bibinfo {author} {\bibfnamefont {M.}~\bibnamefont {{Andr{\'e}}}},\ and\
  \bibinfo {author} {\bibfnamefont {L.}~\bibnamefont {{Matson}}},\ }\href
  {https://doi.org/10.1029/JA095iA05p05983} {\bibfield  {journal} {\bibinfo
  {journal} {Journal of Geophysical Research}\ }\textbf {\bibinfo {volume}
  {95}},\ \bibinfo {pages} {5983} (\bibinfo {year} {1990})}\BibitemShut
  {NoStop}%
\bibitem [{\citenamefont {{Pritchett}}\ and\ \citenamefont
  {{Strangeway}}(1985)}]{1985JGR....90.9650P}%
  \BibitemOpen
  \bibfield  {author} {\bibinfo {author} {\bibfnamefont {P.~L.}\ \bibnamefont
  {{Pritchett}}}\ and\ \bibinfo {author} {\bibfnamefont {R.~J.}\ \bibnamefont
  {{Strangeway}}},\ }\href {https://doi.org/10.1029/JA090iA10p09650} {\bibfield
   {journal} {\bibinfo  {journal} {Journal of Geophysical Research}\ }\textbf
  {\bibinfo {volume} {90}},\ \bibinfo {pages} {9650} (\bibinfo {year}
  {1985})}\BibitemShut {NoStop}%
\bibitem [{\citenamefont {{Winglee}}\ and\ \citenamefont
  {{Dulk}}(1986)}]{1986ApJ...310..432W}%
  \BibitemOpen
  \bibfield  {author} {\bibinfo {author} {\bibfnamefont {R.~M.}\ \bibnamefont
  {{Winglee}}}\ and\ \bibinfo {author} {\bibfnamefont {G.~A.}\ \bibnamefont
  {{Dulk}}},\ }\href {https://doi.org/10.1086/164696} {\bibfield  {journal}
  {\bibinfo  {journal} {\apj}\ }\textbf {\bibinfo {volume} {310}},\ \bibinfo
  {pages} {432} (\bibinfo {year} {1986})}\BibitemShut {NoStop}%
\bibitem [{\citenamefont {Lyubarsky}(2021)}]{Lyubarsky2021}%
  \BibitemOpen
  \bibfield  {author} {\bibinfo {author} {\bibfnamefont {Y.}~\bibnamefont
  {Lyubarsky}},\ }\bibfield  {journal} {\bibinfo  {journal} {Universe}\
  }\textbf {\bibinfo {volume} {7}},\ \href
  {https://doi.org/10.3390/universe7030056} {10.3390/universe7030056} (\bibinfo
  {year} {2021})\BibitemShut {NoStop}%
\bibitem [{\citenamefont {{Gallant}}\ \emph {et~al.}(1992)\citenamefont
  {{Gallant}}, \citenamefont {{Hoshino}}, \citenamefont {{Langdon}},
  \citenamefont {{Arons}},\ and\ \citenamefont {{Max}}}]{1992ApJ...391...73G}%
  \BibitemOpen
  \bibfield  {author} {\bibinfo {author} {\bibfnamefont {Y.~A.}\ \bibnamefont
  {{Gallant}}}, \bibinfo {author} {\bibfnamefont {M.}~\bibnamefont
  {{Hoshino}}}, \bibinfo {author} {\bibfnamefont {A.~B.}\ \bibnamefont
  {{Langdon}}}, \bibinfo {author} {\bibfnamefont {J.}~\bibnamefont {{Arons}}},\
  and\ \bibinfo {author} {\bibfnamefont {C.~E.}\ \bibnamefont {{Max}}},\ }\href
  {https://doi.org/10.1086/171326} {\bibfield  {journal} {\bibinfo  {journal}
  {Astrophysical Jounral}\ }\textbf {\bibinfo {volume} {391}},\ \bibinfo
  {pages} {73} (\bibinfo {year} {1992})}\BibitemShut {NoStop}%
\bibitem [{\citenamefont {{Amato}}\ and\ \citenamefont
  {{Arons}}(2006)}]{2006ApJ...653..325A}%
  \BibitemOpen
  \bibfield  {author} {\bibinfo {author} {\bibfnamefont {E.}~\bibnamefont
  {{Amato}}}\ and\ \bibinfo {author} {\bibfnamefont {J.}~\bibnamefont
  {{Arons}}},\ }\href {https://doi.org/10.1086/508050} {\bibfield  {journal}
  {\bibinfo  {journal} {\apj}\ }\textbf {\bibinfo {volume} {653}},\ \bibinfo
  {pages} {325} (\bibinfo {year} {2006})},\ \Eprint
  {https://arxiv.org/abs/astro-ph/0609034} {arXiv:astro-ph/0609034 [astro-ph]}
  \BibitemShut {NoStop}%
\bibitem [{\citenamefont {Plotnikov}\ and\ \citenamefont
  {Sironi}(2019)}]{Plotnikov2019}%
  \BibitemOpen
  \bibfield  {author} {\bibinfo {author} {\bibfnamefont {I.}~\bibnamefont
  {Plotnikov}}\ and\ \bibinfo {author} {\bibfnamefont {L.}~\bibnamefont
  {Sironi}},\ }\bibfield  {journal} {\bibinfo  {journal} {Monthly Notices of
  the Royal Astronomical Society}\ }\textbf {\bibinfo {volume} {485}},\ \href
  {https://doi.org/10.1093/mnras/stz640} {10.1093/mnras/stz640} (\bibinfo
  {year} {2019})\BibitemShut {NoStop}%
\bibitem [{\citenamefont {{Alsop}}\ and\ \citenamefont
  {{Arons}}(1988)}]{1988PhFl...31..839A}%
  \BibitemOpen
  \bibfield  {author} {\bibinfo {author} {\bibfnamefont {D.}~\bibnamefont
  {{Alsop}}}\ and\ \bibinfo {author} {\bibfnamefont {J.}~\bibnamefont
  {{Arons}}},\ }\href {https://doi.org/10.1063/1.866765} {\bibfield  {journal}
  {\bibinfo  {journal} {Physics of Fluids}\ }\textbf {\bibinfo {volume} {31}},\
  \bibinfo {pages} {839} (\bibinfo {year} {1988})}\BibitemShut {NoStop}%
\bibitem [{\citenamefont {{Hoshino}}\ and\ \citenamefont
  {{Arons}}(1991)}]{1991PhFlB...3..818H}%
  \BibitemOpen
  \bibfield  {author} {\bibinfo {author} {\bibfnamefont {M.}~\bibnamefont
  {{Hoshino}}}\ and\ \bibinfo {author} {\bibfnamefont {J.}~\bibnamefont
  {{Arons}}},\ }\href {https://doi.org/10.1063/1.859877} {\bibfield  {journal}
  {\bibinfo  {journal} {Physics of Fluids B}\ }\textbf {\bibinfo {volume}
  {3}},\ \bibinfo {pages} {818} (\bibinfo {year} {1991})}\BibitemShut {NoStop}%
\bibitem [{\citenamefont {{Iwamoto}}\ \emph {et~al.}(2017)\citenamefont
  {{Iwamoto}}, \citenamefont {{Amano}}, \citenamefont {{Hoshino}},\ and\
  \citenamefont {{Matsumoto}}}]{2017ApJ...840...52I}%
  \BibitemOpen
  \bibfield  {author} {\bibinfo {author} {\bibfnamefont {M.}~\bibnamefont
  {{Iwamoto}}}, \bibinfo {author} {\bibfnamefont {T.}~\bibnamefont {{Amano}}},
  \bibinfo {author} {\bibfnamefont {M.}~\bibnamefont {{Hoshino}}},\ and\
  \bibinfo {author} {\bibfnamefont {Y.}~\bibnamefont {{Matsumoto}}},\ }\href
  {https://doi.org/10.3847/1538-4357/aa6d6f} {\bibfield  {journal} {\bibinfo
  {journal} {Astrophysical Journal}\ }\textbf {\bibinfo {volume} {840}},\
  \bibinfo {eid} {52} (\bibinfo {year} {2017})},\ \Eprint
  {https://arxiv.org/abs/1704.04411} {arXiv:1704.04411 [astro-ph.HE]}
  \BibitemShut {NoStop}%
\bibitem [{\citenamefont {Wu}\ and\ \citenamefont {Yoon}(2007)}]{Wu2007}%
  \BibitemOpen
  \bibfield  {author} {\bibinfo {author} {\bibfnamefont {C.~S.}\ \bibnamefont
  {Wu}}\ and\ \bibinfo {author} {\bibfnamefont {P.~H.}\ \bibnamefont {Yoon}},\
  }\bibfield  {journal} {\bibinfo  {journal} {Physical Review Letters}\
  }\textbf {\bibinfo {volume} {99}},\ \href
  {https://doi.org/10.1103/PhysRevLett.99.075001}
  {10.1103/PhysRevLett.99.075001} (\bibinfo {year} {2007})\BibitemShut
  {NoStop}%
\bibitem [{\citenamefont {Yoon}\ \emph {et~al.}(2009)\citenamefont {Yoon},
  \citenamefont {Wang},\ and\ \citenamefont {Wu}}]{Yoon2009}%
  \BibitemOpen
  \bibfield  {author} {\bibinfo {author} {\bibfnamefont {P.~H.}\ \bibnamefont
  {Yoon}}, \bibinfo {author} {\bibfnamefont {C.~B.}\ \bibnamefont {Wang}},\
  and\ \bibinfo {author} {\bibfnamefont {C.~S.}\ \bibnamefont {Wu}},\
  }\bibfield  {journal} {\bibinfo  {journal} {Physics of Plasmas}\ }\textbf
  {\bibinfo {volume} {16}},\ \href {https://doi.org/10.1063/1.3236749}
  {10.1063/1.3236749} (\bibinfo {year} {2009})\BibitemShut {NoStop}%
\bibitem [{\citenamefont {Zhao}\ and\ \citenamefont {Wu}(2013)}]{Zhao2013}%
  \BibitemOpen
  \bibfield  {author} {\bibinfo {author} {\bibfnamefont {G.~Q.}\ \bibnamefont
  {Zhao}}\ and\ \bibinfo {author} {\bibfnamefont {C.~S.}\ \bibnamefont {Wu}},\
  }\bibfield  {journal} {\bibinfo  {journal} {Physics of Plasmas}\ }\textbf
  {\bibinfo {volume} {20}},\ \href {https://doi.org/10.1063/1.4798493}
  {10.1063/1.4798493} (\bibinfo {year} {2013})\BibitemShut {NoStop}%
\bibitem [{\citenamefont {Wu}\ \emph {et~al.}(2014)\citenamefont {Wu},
  \citenamefont {Chen}, \citenamefont {Zhao},\ and\ \citenamefont
  {Tang}}]{Wu2014}%
  \BibitemOpen
  \bibfield  {author} {\bibinfo {author} {\bibfnamefont {D.~J.}\ \bibnamefont
  {Wu}}, \bibinfo {author} {\bibfnamefont {L.}~\bibnamefont {Chen}}, \bibinfo
  {author} {\bibfnamefont {G.~Q.}\ \bibnamefont {Zhao}},\ and\ \bibinfo
  {author} {\bibfnamefont {J.~F.}\ \bibnamefont {Tang}},\ }\bibfield  {journal}
  {\bibinfo  {journal} {Astronomy and Astrophysics}\ }\textbf {\bibinfo
  {volume} {566}},\ \href {https://doi.org/10.1051/0004-6361/201423898}
  {10.1051/0004-6361/201423898} (\bibinfo {year} {2014})\BibitemShut {NoStop}%
\bibitem [{\citenamefont {Vedenov}\ \emph {et~al.}(1961)\citenamefont
  {Vedenov}, \citenamefont {Velikhov},\ and\ \citenamefont
  {Sagdeev}}]{vedenov1961nonlinear}%
  \BibitemOpen
  \bibfield  {author} {\bibinfo {author} {\bibfnamefont {A.}~\bibnamefont
  {Vedenov}}, \bibinfo {author} {\bibfnamefont {E.}~\bibnamefont {Velikhov}},\
  and\ \bibinfo {author} {\bibfnamefont {R.}~\bibnamefont {Sagdeev}},\
  }\href@noop {} {\bibfield  {journal} {\bibinfo  {journal} {Nuclear Fusion}\
  }\textbf {\bibinfo {volume} {1}},\ \bibinfo {pages} {82} (\bibinfo {year}
  {1961})}\BibitemShut {NoStop}%
\bibitem [{\citenamefont {Drummond}\ and\ \citenamefont
  {Rosenbluth}(1962)}]{drummond1962anomalous}%
  \BibitemOpen
  \bibfield  {author} {\bibinfo {author} {\bibfnamefont {W.~E.}\ \bibnamefont
  {Drummond}}\ and\ \bibinfo {author} {\bibfnamefont {M.~N.}\ \bibnamefont
  {Rosenbluth}},\ }\href@noop {} {\bibfield  {journal} {\bibinfo  {journal}
  {The Physics of Fluids}\ }\textbf {\bibinfo {volume} {5}},\ \bibinfo {pages}
  {1507} (\bibinfo {year} {1962})}\BibitemShut {NoStop}%
\bibitem [{\citenamefont {Stix}(1992)}]{stix1992waves}%
  \BibitemOpen
  \bibfield  {author} {\bibinfo {author} {\bibfnamefont {T.~H.}\ \bibnamefont
  {Stix}},\ }\href@noop {} {\emph {\bibinfo {title} {Waves in plasmas}}}\
  (\bibinfo  {publisher} {Springer Science \& Business Media},\ \bibinfo {year}
  {1992})\BibitemShut {NoStop}%
\bibitem [{\citenamefont {Asenjo}\ \emph {et~al.}(2009)\citenamefont {Asenjo},
  \citenamefont {Muoz}, \citenamefont {Valdivia},\ and\ \citenamefont
  {Hada}}]{Asenjo2009}%
  \BibitemOpen
  \bibfield  {author} {\bibinfo {author} {\bibfnamefont {F.~A.}\ \bibnamefont
  {Asenjo}}, \bibinfo {author} {\bibfnamefont {V.}~\bibnamefont {Muoz}},
  \bibinfo {author} {\bibfnamefont {J.~A.}\ \bibnamefont {Valdivia}},\ and\
  \bibinfo {author} {\bibfnamefont {T.}~\bibnamefont {Hada}},\ }\bibfield
  {journal} {\bibinfo  {journal} {Physics of Plasmas}\ }\textbf {\bibinfo
  {volume} {16}},\ \href {https://doi.org/10.1063/1.3272667}
  {10.1063/1.3272667} (\bibinfo {year} {2009})\BibitemShut {NoStop}%
\bibitem [{\citenamefont {Muñoz}\ \emph {et~al.}(2014)\citenamefont {Muñoz},
  \citenamefont {Asenjo}, \citenamefont {Domínguez}, \citenamefont {López},
  \citenamefont {Valdivia}, \citenamefont {Viñas},\ and\ \citenamefont
  {Hada}}]{Munoz}%
  \BibitemOpen
  \bibfield  {author} {\bibinfo {author} {\bibfnamefont {V.}~\bibnamefont
  {Muñoz}}, \bibinfo {author} {\bibfnamefont {F.~A.}\ \bibnamefont {Asenjo}},
  \bibinfo {author} {\bibfnamefont {M.}~\bibnamefont {Domínguez}}, \bibinfo
  {author} {\bibfnamefont {R.~A.}\ \bibnamefont {López}}, \bibinfo {author}
  {\bibfnamefont {J.~A.}\ \bibnamefont {Valdivia}}, \bibinfo {author}
  {\bibfnamefont {A.}~\bibnamefont {Viñas}},\ and\ \bibinfo {author}
  {\bibfnamefont {T.}~\bibnamefont {Hada}},\ }\bibfield  {journal} {\bibinfo
  {journal} {Nonlinear Processes in Geophysics}\ }\textbf {\bibinfo {volume}
  {21}},\ \href {https://doi.org/10.5194/npg-21-217-2014}
  {10.5194/npg-21-217-2014} (\bibinfo {year} {2014})\BibitemShut {NoStop}%
\bibitem [{\citenamefont {Press}\ \emph {et~al.}(2007)\citenamefont {Press},
  \citenamefont {Teukolsky}, \citenamefont {Vettering},\ and\ \citenamefont
  {Flannery}}]{Press2007}%
  \BibitemOpen
  \bibfield  {author} {\bibinfo {author} {\bibfnamefont {W.~H.}\ \bibnamefont
  {Press}}, \bibinfo {author} {\bibfnamefont {S.~A.}\ \bibnamefont
  {Teukolsky}}, \bibinfo {author} {\bibfnamefont {W.~T.}\ \bibnamefont
  {Vettering}},\ and\ \bibinfo {author} {\bibfnamefont {B.~P.}\ \bibnamefont
  {Flannery}},\ }\href {https://doi.org/10.1017/CBO9781107415324.004} {\emph
  {\bibinfo {title} {NUMERICAL RECIPES The Art of Scientific Computing Third
  Edition}}}\ (\bibinfo {year} {2007})\BibitemShut {NoStop}%
\bibitem [{Note1()}]{Note1}%
  \BibitemOpen
  \bibinfo {note} {See Supplemental Material at [URL will be inserted by
  publisher] for figure showing example comparisons}\BibitemShut {NoStop}%
\bibitem [{\citenamefont {Sironi}\ \emph {et~al.}(2021)\citenamefont {Sironi},
  \citenamefont {Plotnikov}, \citenamefont {Nättilä},\ and\ \citenamefont
  {Beloborodov}}]{Sironi2021}%
  \BibitemOpen
  \bibfield  {author} {\bibinfo {author} {\bibfnamefont {L.}~\bibnamefont
  {Sironi}}, \bibinfo {author} {\bibfnamefont {I.}~\bibnamefont {Plotnikov}},
  \bibinfo {author} {\bibfnamefont {J.}~\bibnamefont {Nättilä}},\ and\
  \bibinfo {author} {\bibfnamefont {A.~M.}\ \bibnamefont {Beloborodov}},\
  }\bibfield  {journal} {\bibinfo  {journal} {Physical Review Letters}\
  }\textbf {\bibinfo {volume} {127}},\ \href
  {https://doi.org/10.1103/PhysRevLett.127.035101}
  {10.1103/PhysRevLett.127.035101} (\bibinfo {year} {2021})\BibitemShut
  {NoStop}%
\bibitem [{\citenamefont {{Babul}}\ and\ \citenamefont
  {{Sironi}}(2020)}]{2020MNRAS.499.2884B}%
  \BibitemOpen
  \bibfield  {author} {\bibinfo {author} {\bibfnamefont {A.-N.}\ \bibnamefont
  {{Babul}}}\ and\ \bibinfo {author} {\bibfnamefont {L.}~\bibnamefont
  {{Sironi}}},\ }\href {https://doi.org/10.1093/mnras/staa2612} {\bibfield
  {journal} {\bibinfo  {journal} {Monthly Notices of the Royal
  Astronomical Society}\ }\textbf {\bibinfo {volume} {499}},\
  \bibinfo {pages} {2884} (\bibinfo {year} {2020})},\ \Eprint
  {https://arxiv.org/abs/2006.03081} {arXiv:2006.03081 [astro-ph.HE]}
  \BibitemShut {NoStop}%
\bibitem [{\citenamefont {Gary}\ and\ \citenamefont
  {Gary}(1993)}]{gary1993theory}%
  \BibitemOpen
  \bibfield  {author} {\bibinfo {author} {\bibfnamefont {S.~P.}\ \bibnamefont
  {Gary}}\ and\ \bibinfo {author} {\bibfnamefont {S.~P.}\ \bibnamefont
  {Gary}},\ }\href@noop {} {\emph {\bibinfo {title} {Theory of space plasma
  microinstabilities}}},\ \bibinfo {number} {7}\ (\bibinfo  {publisher}
  {Cambridge university press},\ \bibinfo {year} {1993})\BibitemShut {NoStop}%
\bibitem [{\citenamefont {{N{\"a}ttil{\"a}}}\ and\ \citenamefont
  {{Beloborodov}}(2022)}]{2022PhRvL.128g5101N}%
  \BibitemOpen
  \bibfield  {author} {\bibinfo {author} {\bibfnamefont {J.}~\bibnamefont
  {{N{\"a}ttil{\"a}}}}\ and\ \bibinfo {author} {\bibfnamefont {A.~M.}\
  \bibnamefont {{Beloborodov}}},\ }\href
  {https://doi.org/10.1103/PhysRevLett.128.075101} {\bibfield  {journal}
  {\bibinfo  {journal} {\prl}\ }\textbf {\bibinfo {volume} {128}},\ \bibinfo
  {eid} {075101} (\bibinfo {year} {2022})},\ \Eprint
  {https://arxiv.org/abs/2111.15578} {arXiv:2111.15578 [astro-ph.HE]}
  \BibitemShut {NoStop}%
\bibitem [{\citenamefont {{Fedorova}}\ and\ \citenamefont
  {{Rodin}}(2019)}]{2019ARep...63...39F}%
  \BibitemOpen
  \bibfield  {author} {\bibinfo {author} {\bibfnamefont {V.~A.}\ \bibnamefont
  {{Fedorova}}}\ and\ \bibinfo {author} {\bibfnamefont {A.~E.}\ \bibnamefont
  {{Rodin}}},\ }\href {https://doi.org/10.1134/S1063772919010037} {\bibfield
  {journal} {\bibinfo  {journal} {Astronomy Reports}\ }\textbf {\bibinfo
  {volume} {63}},\ \bibinfo {pages} {39} (\bibinfo {year} {2019})},\ \Eprint
  {https://arxiv.org/abs/1812.10716} {arXiv:1812.10716 [astro-ph.IM]}
  \BibitemShut {NoStop}%
\bibitem [{\citenamefont {{Gajjar}}\ \emph {et~al.}(2018)\citenamefont
  {{Gajjar}}, \citenamefont {{Siemion}}, \citenamefont {{Price}}, \citenamefont
  {{Law}}, \citenamefont {{Michilli}}, \citenamefont {{Hessels}}, \citenamefont
  {{Chatterjee}}, \citenamefont {{Archibald}}, \citenamefont {{Bower}},
  \citenamefont {{Brinkman}}, \citenamefont {{Burke-Spolaor}}, \citenamefont
  {{Cordes}}, \citenamefont {{Croft}}, \citenamefont {{Enriquez}},
  \citenamefont {{Foster}}, \citenamefont {{Gizani}}, \citenamefont
  {{Hellbourg}}, \citenamefont {{Isaacson}}, \citenamefont {{Kaspi}},
  \citenamefont {{Lazio}}, \citenamefont {{Lebofsky}}, \citenamefont {{Lynch}},
  \citenamefont {{MacMahon}}, \citenamefont {{McLaughlin}}, \citenamefont
  {{Ransom}}, \citenamefont {{Scholz}}, \citenamefont {{Seymour}},
  \citenamefont {{Spitler}}, \citenamefont {{Tendulkar}}, \citenamefont
  {{Werthimer}},\ and\ \citenamefont {{Zhang}}}]{2018ApJ...863....2G}%
  \BibitemOpen
  \bibfield  {author} {\bibinfo {author} {\bibfnamefont {V.}~\bibnamefont
  {{Gajjar}}}, \bibinfo {author} {\bibfnamefont {A.~P.~V.}\ \bibnamefont
  {{Siemion}}}, \bibinfo {author} {\bibfnamefont {D.~C.}\ \bibnamefont
  {{Price}}}, \bibinfo {author} {\bibfnamefont {C.~J.}\ \bibnamefont {{Law}}},
  \bibinfo {author} {\bibfnamefont {D.}~\bibnamefont {{Michilli}}}, \bibinfo
  {author} {\bibfnamefont {J.~W.~T.}\ \bibnamefont {{Hessels}}}, \bibinfo
  {author} {\bibfnamefont {S.}~\bibnamefont {{Chatterjee}}}, \bibinfo {author}
  {\bibfnamefont {A.~M.}\ \bibnamefont {{Archibald}}}, \bibinfo {author}
  {\bibfnamefont {G.~C.}\ \bibnamefont {{Bower}}}, \bibinfo {author}
  {\bibfnamefont {C.}~\bibnamefont {{Brinkman}}}, \bibinfo {author}
  {\bibfnamefont {S.}~\bibnamefont {{Burke-Spolaor}}}, \bibinfo {author}
  {\bibfnamefont {J.~M.}\ \bibnamefont {{Cordes}}}, \bibinfo {author}
  {\bibfnamefont {S.}~\bibnamefont {{Croft}}}, \bibinfo {author} {\bibfnamefont
  {J.~E.}\ \bibnamefont {{Enriquez}}}, \bibinfo {author} {\bibfnamefont
  {G.}~\bibnamefont {{Foster}}}, \bibinfo {author} {\bibfnamefont
  {N.}~\bibnamefont {{Gizani}}}, \bibinfo {author} {\bibfnamefont
  {G.}~\bibnamefont {{Hellbourg}}}, \bibinfo {author} {\bibfnamefont
  {H.}~\bibnamefont {{Isaacson}}}, \bibinfo {author} {\bibfnamefont {V.~M.}\
  \bibnamefont {{Kaspi}}}, \bibinfo {author} {\bibfnamefont {T.~J.~W.}\
  \bibnamefont {{Lazio}}}, \bibinfo {author} {\bibfnamefont {M.}~\bibnamefont
  {{Lebofsky}}}, \bibinfo {author} {\bibfnamefont {R.~S.}\ \bibnamefont
  {{Lynch}}}, \bibinfo {author} {\bibfnamefont {D.}~\bibnamefont {{MacMahon}}},
  \bibinfo {author} {\bibfnamefont {M.~A.}\ \bibnamefont {{McLaughlin}}},
  \bibinfo {author} {\bibfnamefont {S.~M.}\ \bibnamefont {{Ransom}}}, \bibinfo
  {author} {\bibfnamefont {P.}~\bibnamefont {{Scholz}}}, \bibinfo {author}
  {\bibfnamefont {A.}~\bibnamefont {{Seymour}}}, \bibinfo {author}
  {\bibfnamefont {L.~G.}\ \bibnamefont {{Spitler}}}, \bibinfo {author}
  {\bibfnamefont {S.~P.}\ \bibnamefont {{Tendulkar}}}, \bibinfo {author}
  {\bibfnamefont {D.}~\bibnamefont {{Werthimer}}},\ and\ \bibinfo {author}
  {\bibfnamefont {Y.~G.}\ \bibnamefont {{Zhang}}},\ }\href
  {https://doi.org/10.3847/1538-4357/aad005} {\bibfield  {journal} {\bibinfo
  {journal} {\apj}\ }\textbf {\bibinfo {volume} {863}},\ \bibinfo {eid} {2}
  (\bibinfo {year} {2018})},\ \Eprint {https://arxiv.org/abs/1804.04101}
  {arXiv:1804.04101 [astro-ph.HE]} \BibitemShut {NoStop}%
\bibitem [{\citenamefont {{CHIME/FRB Collaboration}}\ \emph
  {et~al.}(2020)\citenamefont {{CHIME/FRB Collaboration}}, \citenamefont
  {{Andersen}}, \citenamefont {{Bandura}}, \citenamefont {{Bhardwaj}},
  \citenamefont {{Bij}}, \citenamefont {{Boyce}}, \citenamefont {{Boyle}},
  \citenamefont {{Brar}}, \citenamefont {{Cassanelli}}, \citenamefont
  {{Chawla}}, \citenamefont {{Chen}}, \citenamefont {{Cliche}}, \citenamefont
  {{Cook}}, \citenamefont {{Cubranic}}, \citenamefont {{Curtin}}, \citenamefont
  {{Denman}}, \citenamefont {{Dobbs}}, \citenamefont {{Dong}}, \citenamefont
  {{Fandino}}, \citenamefont {{Fonseca}}, \citenamefont {{Gaensler}},
  \citenamefont {{Giri}}, \citenamefont {{Good}}, \citenamefont {{Halpern}},
  \citenamefont {{Hill}}, \citenamefont {{Hinshaw}}, \citenamefont
  {{H{\"o}fer}}, \citenamefont {{Josephy}}, \citenamefont {{Kania}},
  \citenamefont {{Kaspi}}, \citenamefont {{Landecker}}, \citenamefont
  {{Leung}}, \citenamefont {{Li}}, \citenamefont {{Lin}}, \citenamefont
  {{Masui}}, \citenamefont {{McKinven}}, \citenamefont {{Mena-Parra}},
  \citenamefont {{Merryfield}}, \citenamefont {{Meyers}}, \citenamefont
  {{Michilli}}, \citenamefont {{Milutinovic}}, \citenamefont {{Mirhosseini}},
  \citenamefont {{M{\"u}nchmeyer}}, \citenamefont {{Naidu}}, \citenamefont
  {{Newburgh}}, \citenamefont {{Ng}}, \citenamefont {{Patel}}, \citenamefont
  {{Pen}}, \citenamefont {{Pinsonneault-Marotte}}, \citenamefont {{Pleunis}},
  \citenamefont {{Quine}}, \citenamefont {{Rafiei-Ravandi}}, \citenamefont
  {{Rahman}}, \citenamefont {{Ransom}}, \citenamefont {{Renard}}, \citenamefont
  {{Sanghavi}}, \citenamefont {{Scholz}}, \citenamefont {{Shaw}}, \citenamefont
  {{Shin}}, \citenamefont {{Siegel}}, \citenamefont {{Singh}}, \citenamefont
  {{Smegal}}, \citenamefont {{Smith}}, \citenamefont {{Stairs}}, \citenamefont
  {{Tan}}, \citenamefont {{Tendulkar}}, \citenamefont {{Tretyakov}},
  \citenamefont {{Vanderlinde}}, \citenamefont {{Wang}}, \citenamefont
  {{Wulf}},\ and\ \citenamefont {{Zwaniga}}}]{2020Natur.587...54C}%
  \BibitemOpen
  \bibfield  {author} {\bibinfo {author} {\bibnamefont {{CHIME/FRB
  Collaboration}}}, \bibinfo {author} {\bibfnamefont {B.~C.}\ \bibnamefont
  {{Andersen}}}, \bibinfo {author} {\bibfnamefont {K.~M.}\ \bibnamefont
  {{Bandura}}}, \bibinfo {author} {\bibfnamefont {M.}~\bibnamefont
  {{Bhardwaj}}}, \bibinfo {author} {\bibfnamefont {A.}~\bibnamefont {{Bij}}},
  \bibinfo {author} {\bibfnamefont {M.~M.}\ \bibnamefont {{Boyce}}}, \bibinfo
  {author} {\bibfnamefont {P.~J.}\ \bibnamefont {{Boyle}}}, \bibinfo {author}
  {\bibfnamefont {C.}~\bibnamefont {{Brar}}}, \bibinfo {author} {\bibfnamefont
  {T.}~\bibnamefont {{Cassanelli}}}, \bibinfo {author} {\bibfnamefont
  {P.}~\bibnamefont {{Chawla}}}, \bibinfo {author} {\bibfnamefont
  {T.}~\bibnamefont {{Chen}}}, \bibinfo {author} {\bibfnamefont {J.~F.}\
  \bibnamefont {{Cliche}}}, \bibinfo {author} {\bibfnamefont {A.}~\bibnamefont
  {{Cook}}}, \bibinfo {author} {\bibfnamefont {D.}~\bibnamefont {{Cubranic}}},
  \bibinfo {author} {\bibfnamefont {A.~P.}\ \bibnamefont {{Curtin}}}, \bibinfo
  {author} {\bibfnamefont {N.~T.}\ \bibnamefont {{Denman}}}, \bibinfo {author}
  {\bibfnamefont {M.}~\bibnamefont {{Dobbs}}}, \bibinfo {author} {\bibfnamefont
  {F.~Q.}\ \bibnamefont {{Dong}}}, \bibinfo {author} {\bibfnamefont
  {M.}~\bibnamefont {{Fandino}}}, \bibinfo {author} {\bibfnamefont
  {E.}~\bibnamefont {{Fonseca}}}, \bibinfo {author} {\bibfnamefont {B.~M.}\
  \bibnamefont {{Gaensler}}}, \bibinfo {author} {\bibfnamefont
  {U.}~\bibnamefont {{Giri}}}, \bibinfo {author} {\bibfnamefont {D.~C.}\
  \bibnamefont {{Good}}}, \bibinfo {author} {\bibfnamefont {M.}~\bibnamefont
  {{Halpern}}}, \bibinfo {author} {\bibfnamefont {A.~S.}\ \bibnamefont
  {{Hill}}}, \bibinfo {author} {\bibfnamefont {G.~F.}\ \bibnamefont
  {{Hinshaw}}}, \bibinfo {author} {\bibfnamefont {C.}~\bibnamefont
  {{H{\"o}fer}}}, \bibinfo {author} {\bibfnamefont {A.}~\bibnamefont
  {{Josephy}}}, \bibinfo {author} {\bibfnamefont {J.~W.}\ \bibnamefont
  {{Kania}}}, \bibinfo {author} {\bibfnamefont {V.~M.}\ \bibnamefont
  {{Kaspi}}}, \bibinfo {author} {\bibfnamefont {T.~L.}\ \bibnamefont
  {{Landecker}}}, \bibinfo {author} {\bibfnamefont {C.}~\bibnamefont
  {{Leung}}}, \bibinfo {author} {\bibfnamefont {D.~Z.}\ \bibnamefont {{Li}}},
  \bibinfo {author} {\bibfnamefont {H.~H.}\ \bibnamefont {{Lin}}}, \bibinfo
  {author} {\bibfnamefont {K.~W.}\ \bibnamefont {{Masui}}}, \bibinfo {author}
  {\bibfnamefont {R.}~\bibnamefont {{McKinven}}}, \bibinfo {author}
  {\bibfnamefont {J.}~\bibnamefont {{Mena-Parra}}}, \bibinfo {author}
  {\bibfnamefont {M.}~\bibnamefont {{Merryfield}}}, \bibinfo {author}
  {\bibfnamefont {B.~W.}\ \bibnamefont {{Meyers}}}, \bibinfo {author}
  {\bibfnamefont {D.}~\bibnamefont {{Michilli}}}, \bibinfo {author}
  {\bibfnamefont {N.}~\bibnamefont {{Milutinovic}}}, \bibinfo {author}
  {\bibfnamefont {A.}~\bibnamefont {{Mirhosseini}}}, \bibinfo {author}
  {\bibfnamefont {M.}~\bibnamefont {{M{\"u}nchmeyer}}}, \bibinfo {author}
  {\bibfnamefont {A.}~\bibnamefont {{Naidu}}}, \bibinfo {author} {\bibfnamefont
  {L.~B.}\ \bibnamefont {{Newburgh}}}, \bibinfo {author} {\bibfnamefont
  {C.}~\bibnamefont {{Ng}}}, \bibinfo {author} {\bibfnamefont {C.}~\bibnamefont
  {{Patel}}}, \bibinfo {author} {\bibfnamefont {U.~L.}\ \bibnamefont {{Pen}}},
  \bibinfo {author} {\bibfnamefont {T.}~\bibnamefont {{Pinsonneault-Marotte}}},
  \bibinfo {author} {\bibfnamefont {Z.}~\bibnamefont {{Pleunis}}}, \bibinfo
  {author} {\bibfnamefont {B.~M.}\ \bibnamefont {{Quine}}}, \bibinfo {author}
  {\bibfnamefont {M.}~\bibnamefont {{Rafiei-Ravandi}}}, \bibinfo {author}
  {\bibfnamefont {M.}~\bibnamefont {{Rahman}}}, \bibinfo {author}
  {\bibfnamefont {S.~M.}\ \bibnamefont {{Ransom}}}, \bibinfo {author}
  {\bibfnamefont {A.}~\bibnamefont {{Renard}}}, \bibinfo {author}
  {\bibfnamefont {P.}~\bibnamefont {{Sanghavi}}}, \bibinfo {author}
  {\bibfnamefont {P.}~\bibnamefont {{Scholz}}}, \bibinfo {author}
  {\bibfnamefont {J.~R.}\ \bibnamefont {{Shaw}}}, \bibinfo {author}
  {\bibfnamefont {K.}~\bibnamefont {{Shin}}}, \bibinfo {author} {\bibfnamefont
  {S.~R.}\ \bibnamefont {{Siegel}}}, \bibinfo {author} {\bibfnamefont
  {S.}~\bibnamefont {{Singh}}}, \bibinfo {author} {\bibfnamefont {R.~J.}\
  \bibnamefont {{Smegal}}}, \bibinfo {author} {\bibfnamefont {K.~M.}\
  \bibnamefont {{Smith}}}, \bibinfo {author} {\bibfnamefont {I.~H.}\
  \bibnamefont {{Stairs}}}, \bibinfo {author} {\bibfnamefont {C.~M.}\
  \bibnamefont {{Tan}}}, \bibinfo {author} {\bibfnamefont {S.~P.}\ \bibnamefont
  {{Tendulkar}}}, \bibinfo {author} {\bibfnamefont {I.}~\bibnamefont
  {{Tretyakov}}}, \bibinfo {author} {\bibfnamefont {K.}~\bibnamefont
  {{Vanderlinde}}}, \bibinfo {author} {\bibfnamefont {H.}~\bibnamefont
  {{Wang}}}, \bibinfo {author} {\bibfnamefont {D.}~\bibnamefont {{Wulf}}},\
  and\ \bibinfo {author} {\bibfnamefont {A.~V.}\ \bibnamefont {{Zwaniga}}},\
  }\href {https://doi.org/10.1038/s41586-020-2863-y} {\bibfield  {journal}
  {\bibinfo  {journal} {\nat}\ }\textbf {\bibinfo {volume} {587}},\ \bibinfo
  {pages} {54} (\bibinfo {year} {2020})},\ \Eprint
  {https://arxiv.org/abs/2005.10324} {arXiv:2005.10324 [astro-ph.HE]}
  \BibitemShut {NoStop}%
\bibitem [{\citenamefont {{Mereghetti}}(2008)}]{2008A&ARv..15..225M}%
  \BibitemOpen
  \bibfield  {author} {\bibinfo {author} {\bibfnamefont {S.}~\bibnamefont
  {{Mereghetti}}},\ }\href {https://doi.org/10.1007/s00159-008-0011-z}
  {\bibfield  {journal} {\bibinfo  {journal} {Astronomy and Astrophysics
  Review}\ }\textbf {\bibinfo {volume} {15}},\ \bibinfo {pages} {225} (\bibinfo
  {year} {2008})},\ \Eprint {https://arxiv.org/abs/0804.0250} {arXiv:0804.0250
  [astro-ph]} \BibitemShut {NoStop}%
\bibitem [{\citenamefont {{Olausen}}\ and\ \citenamefont
  {{Kaspi}}(2014)}]{2014ApJS..212....6O}%
  \BibitemOpen
  \bibfield  {author} {\bibinfo {author} {\bibfnamefont {S.~A.}\ \bibnamefont
  {{Olausen}}}\ and\ \bibinfo {author} {\bibfnamefont {V.~M.}\ \bibnamefont
  {{Kaspi}}},\ }\href {https://doi.org/10.1088/0067-0049/212/1/6} {\bibfield
  {journal} {\bibinfo  {journal} {Astrophysical Journal Supplement}\ }\textbf
  {\bibinfo {volume} {212}},\ \bibinfo {eid} {6} (\bibinfo {year} {2014})},\
  \Eprint {https://arxiv.org/abs/1309.4167} {arXiv:1309.4167 [astro-ph.HE]}
  \BibitemShut {NoStop}%
\bibitem [{Note2()}]{Note2}%
  \BibitemOpen
  \bibinfo {note}
  {Http://www.physics.mcgill.ca/~pulsar/magnetar/main.html}\BibitemShut
  {NoStop}%
\bibitem [{\citenamefont {{Dall'Osso}}\ and\ \citenamefont
  {{Stella}}(2022)}]{2022ASSL..465..245D}%
  \BibitemOpen
  \bibfield  {author} {\bibinfo {author} {\bibfnamefont {S.}~\bibnamefont
  {{Dall'Osso}}}\ and\ \bibinfo {author} {\bibfnamefont {L.}~\bibnamefont
  {{Stella}}},\ }in\ \href {https://doi.org/10.1007/978-3-030-85198-9\_8}
  {\emph {\bibinfo {booktitle} {Astrophysics and Space Science Library}}},\
  \bibinfo {series} {Astrophysics and Space Science Library}, Vol.\ \bibinfo
  {volume} {465},\ \bibinfo {editor} {edited by\ \bibinfo {editor}
  {\bibfnamefont {S.}~\bibnamefont {{Bhattacharyya}}}, \bibinfo {editor}
  {\bibfnamefont {A.}~\bibnamefont {{Papitto}}},\ and\ \bibinfo {editor}
  {\bibfnamefont {D.}~\bibnamefont {{Bhattacharya}}}}\ (\bibinfo {year}
  {2022})\ pp.\ \bibinfo {pages} {245--280},\ \Eprint
  {https://arxiv.org/abs/2103.10878} {arXiv:2103.10878 [astro-ph.HE]}
  \BibitemShut {NoStop}%
\bibitem [{\citenamefont {{Margalit}}\ and\ \citenamefont
  {{Metzger}}(2018)}]{2018ApJ...868L...4M}%
  \BibitemOpen
  \bibfield  {author} {\bibinfo {author} {\bibfnamefont {B.}~\bibnamefont
  {{Margalit}}}\ and\ \bibinfo {author} {\bibfnamefont {B.~D.}\ \bibnamefont
  {{Metzger}}},\ }\href {https://doi.org/10.3847/2041-8213/aaedad} {\bibfield
  {journal} {\bibinfo  {journal} {Astrophysical Journal Letters}\ }\textbf
  {\bibinfo {volume} {868}},\ \bibinfo {eid} {L4} (\bibinfo {year} {2018})},\
  \Eprint {https://arxiv.org/abs/1808.09969} {arXiv:1808.09969 [astro-ph.HE]}
  \BibitemShut {NoStop}%
\end{thebibliography}%
\end{document}